\documentclass[12pt]{iopart}

\usepackage{graphicx}

\begin{document}

\title[Linearized stability analysis of thin-shell wormholes with
$\Lambda$]{Linearized stability analysis of thin-shell wormholes
with a cosmological constant}

\author{Francisco S. N. Lobo\footnote[1]{flobo@cosmo.fis.fc.ul.pt}
 and Paulo Crawford\footnote[2]{crawford@cosmo.fis.fc.ul.pt} }

\address{Centro de Astronomia
e Astrof\'{\i}sica da Universidade de Lisboa,\\
Campo Grande, Ed. C8 1749-016 Lisboa, Portugal}


\begin{abstract}

Spherically symmetric thin-shell wormholes in the presence of a
cosmological constant are constructed applying the cut-and-paste
technique implemented by Visser. Using the Darmois-Israel
formalism the surface stresses, which are concentrated at the
wormhole throat, are determined. This construction allows one to
apply a dynamical analysis to the throat, considering linearized
radial perturbations around static solutions. For a large positive
cosmological constant, i.e., for the Schwarzschild-de Sitter
solution, the region of stability is significantly increased,
relatively to the null cosmological constant case, analyzed by
Poisson and Visser. With a negative cosmological constant, i.e.,
the Schwarzschild-anti de Sitter solution, the region of stability
is decreased. In particular, considering static solutions with a
generic cosmological constant, the weak and dominant energy
conditions are violated, while for $a_0 \leq 3M$ the null and
strong energy conditions are satisfied. The surface pressure of
the static solution is strictly positive for the Schwarzschild and
Schwarzschild-anti de Sitter spacetimes, but takes negative
values, assuming a surface tension in the Schwarzschild-de Sitter
solution, for high values of the cosmological constant and the
wormhole throat radius.

\end{abstract}


\maketitle

\section{Introduction}

Wormhole solutions are necessarily associated with violations of
the energy conditions \cite{Morris}, namely the existence of
exotic matter. Being a problematic issue, it is useful to minimize
the usage of exotic matter, although energy condition violations
in quantum field theory are well known \cite{Morris,Visser}. Thus,
one may concentrate the exotic matter to a thin-shell localized at
the throat, using a simple and elegant class of wormhole solutions
implemented by Matt Visser using the cut-and-paste construction
\cite{Visser,visser1,visserNPB}. The surface stress-energy tensor
components of the exotic matter at the throat are determined,
invoking the Darmois-Israel formalism \cite{Israel, Papahamoui}.

This construction also permits one to explore a dynamical analysis
of the wormhole throat. Poisson and Visser analyzed a thin-shell
wormhole \cite{Poisson}, constructed by pasting together two
copies of the Schwarzschild solution at $a>2M$. Considering a
linearized radial perturbation around a static solution, in the
spirit of \cite{Brady,Balbinot}, regions of stability were
discovered, which lie in a somewhat unexpected patch. This is due
to the fact that one does not have a detailed microphysical model
for exotic matter, in fact, nothing is known on equations of state
of exotic matter. See discussion in Poisson-Visser \cite{Poisson}.
Building on this, Eiroa and Romero \cite{Eiroa} analyzed charged
thin-shell wormholes and found that the regions of stability are
greatly increased for large values of charge.

The presence of a cosmological constant is also an issue that has
been extensively analyzed in the literature. For instance, Kim
\cite{Kim} analyzed Schwarzschild-de Sitter wormholes, using the
cut-and-paste construction. Considering specific equations of
state the role of the cosmological constant in the classical and
quantum stabilities was studied. This analysis was a natural
extension of the Schwarzschild wormhole solution discussed in
\cite{visserNPB,visserPLB}. Later, Kim and collaborators
\cite{Kim2} discussed the $(2+1)$-dimensional Schwarzschild-de
Sitter wormhole, where once again taking into account specific
equations of state, the respective dynamical stability of this
type of wormhole was discussed. A comparison with the
$(3+1)$-dimensional Schwarzschild-de Sitter wormhole and the
$(2+1)$-dimensional discussed in \cite{Perry} was established. A
partial stability analysis of a $(2+1)$-dimensional wormhole
solution, in the presence of $\Lambda$, was explored by Delgaty
and Mann \cite{Delgaty}, in which the simple case of a constant
line energy density was considered.

More recently, prompted by the Randall-Sundrum brane world
scenario \cite{Randall,Randall2}, where our universe is viewed as
a domain wall in five dimensional anti-de Sitter space, Kraus
\cite{Kraus} analyzed the dynamics of $5$-dimensional anti-de
Sitter domain walls. The motion of the wall was determined using
the Darmois-Israel formalism. The possible wall trajectories were
described and the consequences for localized gravity on the wall
were examined. Anchordoqui \etal \cite{Anch} considered the
dynamics of the simple case of onebranes in $3$-dimensional
anti-de Sitter space and generalized the previous analysis for an
arbitrary number of dimensions, studying the quantum behaviour
within the context of the WKB approximation. Using the standard
linearization analysis, Anchordoqui and Olsen \cite{Anch2} studied
the classical stability and different behaviours for the expansion
of the universe by imposing constraints on the speed of sound,
within $(d+1)$-dimensional anti-de Sitter background spaces. The
formation of black holes due to the gravitational collapse of
matter trapped on the brane, within the context of the
Randall-Sundrum models was discussed in \cite{Chamblin}. Chamblin
\etal conjectured that a non-rotating uncharged black hole on the
domain wall is described by a {\it black cigar} solution in five
dimensions. They also considered the five dimensional
Schwarzschild-anti de Sitter solution as another candidate for a
black hole formed by the gravitational collapse on a domain wall
in anti-de Sitter space. A great deal of attention on the issue of
gravity on the brane within the Randall-Sundrum scenario has also
been devoted to the cosmological solutions
\cite{Garriga,Kaloper,Bin}.

Furthermore, spherically symmetric and static traversable
wormholes in the presence of a generic cosmological constant,
$\Lambda$, were analyzed in \cite{Lemos}. A matching of an
interior solution to the unique exterior vacuum solution was
carried out, and the respective structure and several physical
properties and characteristics due to $\Lambda$ were explored.
Specific cases of thin shells at the matching due to the effects
of the cosmological constant were studied. In the present paper,
we concentrate the exotic matter to the wormhole throat,
generalizing the Poisson-Visser analysis for thin-shell wormholes
in the presence of a non-vanishing cosmological constant. As in
\cite{Poisson,Anch2} we shall consider linearized radial
perturbations around a static wormhole solution, which is
complementary to the analysis discussed by Kim \cite{Kim}. The
advantage of this method resides mainly in the fact that one
defines a parametrization of the stability of equilibrium
\cite{Poisson,Ishak}, as not to specify an equation of state on
the boundary surface. This paper is organized as follows. In
section $2$ the cut-and-paste construction and the Darmois-Israel
formalism are briefly presented, and the surface stresses are
deduced. In section $3$, a linearized stability analysis is
studied in the presence of a generic cosmological constant. Some
properties and characteristics of the static solution is also
analyzed. In section $4$ we conclude and in the Appendix we
briefly consider a linearized stability in $(2+1)$-dimensional
Schwarzschild-de Sitter thin-shell wormholes.

\section{Thin-shell wormholes with a cosmological constant}

To generalize the analysis of Poisson-Visser, consider the unique
spherically symmetric vacuum solution in the presence of a
non-vanishing cosmological constant, i.e.,
\begin{eqnarray}
\fl ds^2=-\left(1-\frac{2M}{r}-\frac{\Lambda}{3}r^2
\right)\,dt^2+\left(1-\frac{2M}{r}-\frac{\Lambda}{3}r^2
\right)^{-1}\,dr^2 +r^2\,(d\theta ^2+\sin ^2{\theta}\, d\phi ^2)
\label{metricvacuumlambda}.
\end{eqnarray}
If $\Lambda >0$, the solution is denoted by the Schwarzschild-de
Sitter metric. For $\Lambda <0$, we have the Schwarzschild-anti de
Sitter metric, and of course the specific case of $\Lambda =0$ is
reduced to the Schwarzschild solution. Note that the metric $(1)$
is not asymptotically flat as $r \rightarrow \infty$. Rather, it
is asymptotically de Sitter, if $\Lambda >0$, or asymptotically
anti-de Sitter, if $\Lambda <0$. But, considering low values of
$\Lambda$, the metric is almost flat in the range $M \ll r \ll
1/\sqrt{\Lambda}$. For values below this range, the effects of $M$
dominate, and for values above the range, the effects of $\Lambda$
dominate, as for large values of the radial coordinate the
large-scale structure of the spacetime must be taken into account.

\subsection{The Schwarzschild-de Sitter spacetime, $\Lambda >0$}

If $0<9\Lambda M^2<1$, the factor $f(r)=(1-2M/r-\Lambda r^2/3)$
possesses two positive real roots, $r_b$ and $r_c$, corresponding
to the black hole and the cosmological event horizons of the de
Sitter spacetime, respectively. Considering the following
definitions
\begin{eqnarray}
A&=&\frac{1}{2M}\,\sqrt[3]{\,\frac{3M}{\Lambda}\left(-1+\sqrt{1-
(9\Lambda M^2)^{-1}}\,\right)}
\label{A}   \,, \\
B&=&\frac{1}{2M}\,\sqrt[3]{\,\frac{3M}{\Lambda}\left(-1-\sqrt{1-(9\Lambda
M^2)^{-1}}\, \right)} \label{B}  \,,
\end{eqnarray}
we verify that
\begin{eqnarray}
r_b&=&-M \left[(A+B)+\sqrt{-3}(A-B) \right]
\label{root1}  \,, \\
r_c&=&2M\,(A+B)    \label{root2}  \,,
\end{eqnarray}
are the black hole and the cosmological event horizons,
respectively. Considering $\Lambda M^2 \ll 1$, we have
\begin{eqnarray}
r_b&=&2M \left(1+4 \Lambda M^2/3\right)\,, \\
r_c&=&\sqrt{3/\Lambda} \left(1-M\sqrt{\Lambda/3}\,\right)
\label{cosmologicalhorizon} \,.
\end{eqnarray}
See \cite{Lemos} for details. If $9\Lambda M^2=1$, both horizons
coincide at $r_b=r_c=3M$.

\subsection{The Schwarzschild-anti de Sitter spacetime, $\Lambda
<0$}

For the Schwarzschild-anti de Sitter metric, with $\Lambda <0$,
the factor $f(r)=(1-2M/r+|\Lambda |r^2/3 )$ has only one real
positive root, $r_b$, given by
\begin{eqnarray}
r_b=\left(\frac{3M}{|\Lambda|}\right)^{1/3}\left(\sqrt[3]{1+\sqrt{1+(9|\Lambda|M^2)^{-1}}}
+\sqrt[3]{1-\sqrt{1+ (9|\Lambda|M^2)^{-1}}}\;\right)
\label{adsbhole} ,
\end{eqnarray}
corresponding to a black hole event horizon. For $|\Lambda |M^2
\ll 1$ one obtains
\begin{equation}
r_b=2M \left(1-4 |\Lambda |M^2/3 \right)\,.
\end{equation}
See \cite{Lemos} for details.

\subsection{The Darmois-Israel formalism and the cut-and-paste
construction}

Given this, we may construct a wormhole solution, using the
cut-and-paste technique \cite{Visser}. Consider two vacuum
solutions with $\Lambda$ and remove from each spacetime the region
described by
\begin{equation}
\Omega_{1,2}\equiv \left \{r_{1,2}\leq a| \,a >r_b \right \} \,,
\end{equation}
where $a$ is a constant and $r_b$ is the black hole event horizon,
corresponding to the Schwarzschild-de Sitter and
Schwarzschild-anti de Sitter solutions, equation (\ref{root1}) and
equation (\ref{adsbhole}), respectively. The removal of the
regions results in two geodesically incomplete manifolds, with
boundaries given by the following timelike hypersurfaces
\begin{equation}
\partial \Omega_{1,2}\equiv \left \{r_{1,2}= a| \,a > r_b \right \} \,.
\end{equation}
Identifying these two timelike hypersurfaces, $\partial
\Omega_{1}=\partial \Omega_{2}$, results in a geodesically
complete manifold, with two regions connected by a wormhole and
the respective throat situated at $\partial \Omega$. The wormhole
connects two regions, asymptotically de Sitter or anti-de Sitter,
for $\Lambda >0$ and $\Lambda <0$, respectively.

We shall use the Darmois-Israel formalism to determine the surface
stresses at the junction boundary \cite{Israel, Papahamoui}. The
intrinsic surface stress-energy tensor, $S_{ij}$, is given by the
Lanczos equations \cite{Israel} in the form
$S^{i}_{\;j}=-\frac{1}{8\pi}\,(\kappa ^{i}_{\;\,j}-\delta
^{i}_{\;j}\kappa ^{k}_{\;\,k})$. For notational convenience, the
discontinuity in the second fundamental form or extrinsic
curvatures is given by $\kappa_{ij}=K_{ij}^{+}-K_{ij}^{-}$. The
second fundamental form is defined as
\begin{eqnarray}
K_{ij}^{\pm}&=&\frac{\partial x^{\alpha}}{\partial \xi ^{i}} \,
\frac{\partial x^{\beta}}{\partial \xi
^{j}}\,\nabla_{\alpha}^{\pm}\,n_{\beta}   \nonumber   \\
&=&-n_{\gamma} \left(\frac{\partial ^2 x^{\gamma}}{\partial \xi
^{i}\,\partial \xi ^{j}}+\Gamma ^{\gamma \pm}_{\;\;\alpha
\beta}\;\frac{\partial x^{\alpha}}{\partial \xi ^{i}} \,
\frac{\partial x^{\beta}}{\partial \xi ^{j}} \right)
   \label{defextrinsiccurvature}     \,,
\end{eqnarray}
where $n_{\gamma}$ are the components of the unit normal vector to
$\partial \Omega$, and $\xi ^{i}$ represent the intrinsic
coordinates in $\partial \Omega$. The superscripts $\pm$
correspond to the exterior and interior spacetimes, respectively.
The parametric equation for $\partial \Omega$ is given by
$f(x^{\mu}(\xi ^{i}))=0$ and the respective unit $4$-normal to
$\partial \Omega$ is defined by
\begin{equation}
n_{\mu}=\pm \Bigg| g^{\alpha \beta}\, \frac{\partial f}{\partial
x^{\alpha}}\, \frac{\partial f}{\partial x^{\beta}} \Bigg|^{-1/2}
\;\frac{\partial f}{\partial x^{\mu}} \label{defnormal} \,,
\end{equation}
with $n^{\mu}\,n_{\mu}=+1$. The intrinsic metric at $\partial
\Omega$ is given by
\begin{equation}
ds^2\Big|_{\partial \Omega}=-d\tau ^2 +a^2(\tau)\,(d\theta ^2+\sin
^2{\theta}\, d\phi ^2)\,,
\end{equation}
where $\tau$ is the proper time as measured by a comoving observer
on the wormhole throat.

\subsection{The surface stresses}

Considerable simplifications occur due to spherical symmetry,
namely
\[
\kappa ^{i}_{\;j}={\rm diag} \left(\kappa
^{\tau}_{\;\,\tau},\kappa ^{\theta}_{\;\,\theta},\kappa
^{\theta}_{\;\,\theta}\right).
\]
Thus, the surface stress-energy tensor may be written in terms of
the surface energy density, $\sigma$, and the surface pressure,
$p$, as
\[
S^{i}_{\;j}={\rm diag}(-\sigma,p,p)\,,
\]
which taking into account the Lanczos equations, reduce to
\begin{eqnarray}
\sigma &=&-\frac{1}{4\pi}\kappa ^{\theta}_{\;\,\theta} \,,   \label{sigma} \\
p &=&\frac{1}{8\pi}(\kappa ^{\tau}_{\;\,\tau}+\kappa
^{\theta}_{\;\,\theta})    \label{surfacepressure}  \,.
\end{eqnarray}
This simplifies the determination of the surface stress-energy
tensor to that of the calculation of the non-trivial components of
the extrinsic curvature.

The imposition of spherical symmetry is sufficient to conclude
that there is no gravitational radiation, independently of the
behavior of the wormhole throat. The position of the throat is
given by
$x^{\mu}(\tau,\theta,\phi)=(t(\tau),a(\tau),\theta,\phi)$, and the
respective $4$-velocity is
\begin{equation}
U^{\mu}=\left(\frac{\sqrt{1-2M/a-\Lambda a^2/3+\dot{a}^2}}
{1-2M/a-\Lambda a^2/3},\dot{a},0,0 \right)  \,,
\end{equation}
where the overdot denotes a derivative with respect to $\tau$.

The unit normal to the throat may be determined by equation
(\ref{defnormal}) or by the contractions, $U^{\mu}n_{\mu}=0$ and
$n^{\mu}n_{\mu}=+1$, and is given by
\begin{equation}
n^{\mu}=\left(\frac{\dot{a}} {1-2M/a-\Lambda
a^2/3},\sqrt{1-2M/a-\Lambda a^2/3+\dot{a}^2},0,0 \right)
\label{normal} \,.
\end{equation}

Using equation (\ref{metricvacuumlambda}), equation
(\ref{defextrinsiccurvature}) and equation (\ref{normal}), the
non-trivial components of the extrinsic curvature are given by
\begin{eqnarray}
K^{\theta\;\pm}_{\;\,\theta}&=& \pm
\frac{1}{a}\sqrt{1-2M/a-\Lambda a^2/3+\dot{a}^2} \;,
    \label{Kplustheta}\\
K ^{\tau \;\pm}_{\;\,\tau}&=& \pm \frac{M/a^2-\Lambda a/3+
\ddot{a}} {\sqrt{1-2M/a-\Lambda a^2/3+\dot{a}^2}} \;,
    \label{Kplustautau}
\end{eqnarray}
Thus, the Einstein field equations, equations
(\ref{sigma})-(\ref{surfacepressure}), with equations
(\ref{Kplustheta})-(\ref{Kplustautau}), then provide us with the
following surface stresses
\begin{eqnarray}
\sigma&=&-\frac{1}{2\pi a} \sqrt{1-2M/a-\Lambda a^2/3+\dot{a}^2}
  \label{surfenergy}   \,,\\
p&=&\frac{1}{4\pi a} \;\frac{1-M/a-2\Lambda
a^2/3+\dot{a}^2+a\ddot{a}}{\sqrt{1-2M/a-\Lambda a^2/3+\dot{a}^2}}
\label{surfpressure}  \,.
\end{eqnarray}

We also verify that the above equations imply the conservation of
the surface stress-energy tensor
\begin{equation}
\dot{\sigma}=-2 \left(\sigma +p \right) \frac{\dot{a}}{a}
   \label{conservationenergy}
\end{equation}
or
\begin{equation}
\frac{d\left(\sigma A \right)}{d\tau}+p\,\frac{dA}{d\tau}=0 \,,
\end{equation}
where $A=4\pi a^2$ is the area of the wormhole throat. The first
term represents the variation of the internal energy of the
throat, and the second term is the work done by the throat's
internal forces.

\section{Linearized stability analysis}

Equation (\ref{surfenergy}) may be recast into the following
dynamical form
\begin{equation}
\dot{a}^2-\frac{2M}{a}-\frac{\Lambda}{3} a^2- \left(2\pi \sigma a
\right)^2=-1  \,,  \label{motionofthroat}
\end{equation}
which determines the motion of the wormhole throat. Considering an
equation of state of the form, $p=p(\sigma)$, the energy
conservation, equation (\ref{conservationenergy}), can be
integrated to yield
\begin{equation}
\ln(a)=-\frac{1}{2}\int \frac{d\sigma}{\sigma +p(\sigma)}  \,.
\end{equation}
This result formally inverted to provide $\sigma=\sigma(a)$, can
be finally substituted into equation (\ref{motionofthroat}).
Equation (\ref{motionofthroat}) can also be written as
$\dot{a}^2=-V(a)$, with the potential defined as
\begin{equation}
V(a)=1-\frac{2M}{a}-\frac{\Lambda}{3} a^2- \left(2\pi \sigma a
\right)^2     \label{defpotential}  \,.
\end{equation}

\subsection{Static solution}

One may explore specific equations of state, but following the
Poisson-Visser reasoning \cite{Poisson}, we shall consider a
linear perturbation around a static solution with radius $a_0$.
The respective values of the surface energy density and the
surface pressure, at $a_0$, are given by
\begin{eqnarray}
\sigma_0&=&-\frac{1}{2\pi a_0} \sqrt{1-2M/a_0-\Lambda
a_0^2/3}   \label{stablesigma} \,,\\
p_0&=&\frac{1}{4\pi a_0}\;\frac{1-M/a_0-2\Lambda
a_0^2/3}{\sqrt{1-2M/a_0-\Lambda a_0^2/3}} \,.
\label{stablepressure}
\end{eqnarray}
One verifies that the surface energy density is always negative,
implying the violation of the weak and dominant energy conditions.

The null energy condition is satisfied if $\sigma_0+p_0 \geq 0$.
Taking into account the following relationship
\begin{equation}
\sigma_0+p_0=-\frac{1}{4\pi a_0} \;
\frac{1-3M/a_0}{\sqrt{1-2M/a_0-\Lambda a_0^2/3}} \label{NEC} \,,
\end{equation}
the null energy condition is satisfied for $a_0 \leq 3M$,
considering a generic cosmological constant.

The strong energy condition is satisfied if $\sigma_0+p_0 \geq 0$
and $\sigma_0+2p_0 \geq 0$, and by continuity implies the null
energy condition. Consider
\begin{equation}
\sigma_0+2p_0=\frac{1}{2\pi a_0} \frac{M/a_0-\Lambda
a_0^2/3}{\sqrt{1-2M/a_0-\Lambda a_0^2/3}} \label{SEC} \,.
\end{equation}
For the Schwarzschild and the Schwarzschild-anti de Sitter
spacetimes, $\Lambda \leq 0$ and $M>0$, the condition
$\sigma_0+2p_0 >0$ is verified. The Schwarzschild-de Sitter
solution, $\Lambda >0$, requires a further analysis. Exercising
simple algebraic gymnastics, we verify that the imposition of $a_0
\leq 3M$, so that the null energy condition is satisfied, and
taking into account $9\Lambda M^2 <1$, implies $M/a_0 >\Lambda
a_0^2/3$, so that $\sigma_0+2p_0
>0$. We conclude that the null and strong energy conditions are
satisfied for $a_0 \leq 3M$ for a generic $\Lambda$. In
particular, for $M=0$ and $\Lambda =0$, the above analysis is
reduced to the Minkowski surgery discussed in \cite{Visser}, in
which all the pointwise and averaged energy conditions are
violated for a convex section of the wormhole throat. For $M=0$
and $\Lambda \neq 0$, the de Sitter and the anti-de Sitter
solutions, considering equation (\ref{stablesigma}) and equation
(\ref{NEC}), all the energy conditions are also violated.

An analysis of the sign of $p_0$, equation (\ref{stablepressure}),
is also in order. For the Schwarzschild and the Schwarzschild-anti
de Sitter solutions, $\Lambda \leq 0$, $p_0$ is positive, implying
a surface pressure. But, for the Schwarzschild-de Sitter spacetime
$p_0$ may have a negative value, assuming the character of a
surface tension. The sign of $p_0$ is analyzed in Figure $1$,
where we plot $9\Lambda M^2 \,\times 2M/a_0$. For large values of
the cosmological constant, or for large $M$, and for large values
of $a_0$, one verifies that a surface tension is needed to support
the structure. This is somehow to be expected because for high
values of $\Lambda$ the large-scale curvature of the solution has
to be taken into account. Therefore, one needs a tension to hold
the wormhole from expanding. For low values of $\Lambda$ and for
small $a_0$, $p_0$ is a surface pressure, preventing the wormhole
from collapsing. The qualitative behavior of $p_0$ is similar to
the surface pressure/tension for the thin-shells considered in
\cite{Lemos}.

The behavior of the sign becomes transparent in the following
algebraic analysis. Consider the factors given by
$f(a_0)=(1-2M/a_0-\Lambda a_0^2/3 )$ and $h(a_0)=(1-M/a_0-2\Lambda
a_0^2/3)$. Define $\alpha =2M/a_0$ and $\beta = 9 \Lambda M^2$, so
that $\beta =27\alpha^2(1-\alpha)/4$ is considered for $f(a_0)=0$,
which is depicted as a solid line in Figure $1$. $p_0$ will be
null if $h(a_0)=0$, i.e., for $9\Lambda M^2
=27\alpha^2(1-\alpha/2)/8$, depicted as a dashed line in Figure
$1$. A surface pressure is present, $p_0 >0$, if $h(a_0)>0$, which
is shown to the right of the dashed curve in Figure $1$. A surface
tension, $p_0 <0$, is shown to the left of the dashed line.

\begin{figure}[h]
  \centering
  \includegraphics[width=2.8in]{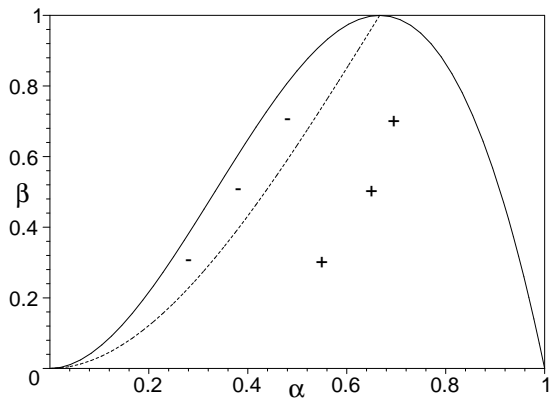}
  \caption{We have defined $\alpha=2M/a_0$ and $\beta=9\Lambda M^2$.
  Only the region below the solid line is considered. The dashed line
  depicts a null surface pressure. To the right of the dashed curve
  we have a surface pressure, and to the left a surface
  tension. See text for details.}
\end{figure}

\subsection{Linearization analysis}

Linearizing around the stable solution at $a=a_0$, we consider a
Taylor expansion of $V(a)$ around $a_0$ to second order, which
provides
\begin{equation}
 V(a)=V(a_0)+V'(a_0)(a-a_0)+\frac{1}{2}\,V''(a_0)(a-a_0)^2+O
\left[(a-a_0)^3 \right] \label{Taylorexpansion} ,
\end{equation}
where the prime denotes a derivative with respect to $a$, $d/da$.

To determine the first and second derivatives of the potential
equation (\ref{defpotential}), $V'(a)$ and $V''(a)$, respectively,
it is useful to rewrite the conservation of the surface
stress-energy tensor, equation (\ref{conservationenergy}), as
$\sigma 'a=-2(\sigma +p)$, taking into account $\sigma
'=\dot{\sigma}/\dot{a}$. Defining the parameter
$\eta(\sigma)=dp/d\sigma=p'/\sigma'$, we have $\sigma '+2p'=\sigma
'(1+2\eta)$. The physical interpretation of $\eta$ is discussed in
\cite{Poisson}, and $\sqrt{\eta}$ is normally interpreted as the
speed of sound. With these results, $V'(a)$ and $V''(a)$ are
respectively given by
\begin{eqnarray}
V'(a)&=&\frac{2M}{a^2}-\frac{2\Lambda}{3}a+8\pi^2\sigma a\,(\sigma
+2p)   \label{firstderivativeV}  \,, \\
V''(a)&=&-\frac{4M}{a^3}-\frac{2\Lambda}{3}-8\pi^2 \big[(\sigma
+2p)^2 \mbox{}  +2\sigma (\sigma +p)(1+2\eta) \big]
\label{2ndderivative} .
\end{eqnarray}
Evaluated at the static solution, at $a=a_0$, using equations
(\ref{stablesigma})-(\ref{stablepressure}) and taking into account
equation (\ref{NEC}) and equation (\ref{SEC}), we readily find
$V(a_0)=0$ and $V'(a_0)=0$, with $V''(a_0)$ given by
\begin{equation}
\fl V''(a_0)=-\frac{2}{a_0^2} \Bigg[
\frac{2M}{a_0}+\frac{\Lambda}{3}a_0^2+\frac{\left(M/a_0-\Lambda
a_0^2/3 \right)^2}{1-2M/a_0-\Lambda a_0^2/3}+(1+2\eta_0)
\left(1-\frac{3M}{a_0} \right) \Bigg],
\end{equation}
where $\eta_0=\eta (\sigma_0)$.

The potential $V(a)$, equation (\ref{Taylorexpansion}), is reduced
to
\begin{equation}
V(a)=\frac{1}{2}\,V''(a_0)(a-a_0)^2+O \left[(a-a_0)^3 \right]  \,,
\end{equation}
so that the equation of motion for the wormhole throat presents
the following form
\begin{equation}
\dot{a}^2=-\frac{1}{2}V''(a_0)(a-a_0)^2+O \left[(a-a_0)^3 \right]
\,,
\end{equation}
to the order of approximation considered. If $V''(a_0)<0$ is
verified, then the potential $V(a_0)$ has a local maximum at
$a_0$, where a small perturbation in the wormhole throat's radius
will provoke an irreversible contraction or expansion of the
throat. Thus, the solution is stable if and only if $V(a_0)$ has a
local minimum at $a_0$ and $V''(a_0)>0$, i.e.,
\begin{equation}
\frac{2M}{a_0}+\frac{\Lambda}{3}a_0^2+\frac{\left(M/a_0-\Lambda
a_0^2/3 \right)^2}{1-2M/a_0-\Lambda a_0^2/3}+\left(1+2\eta_0
\right) \left(1-\frac{3M}{a_0} \right) <0
\end{equation}
or
\begin{equation}
\eta_0 \left(1-\frac{3M}{a_0} \right) <
-\frac{1-3M/a_0+3M^2/a_0^2-\Lambda Ma_0}{2 \left(1-2M/a_0-\Lambda
a_0^2/3 \right)}
   \label{inequality}   \,.
\end{equation}

We need to analyze equation (\ref{inequality}) for several cases.
Firstly, for the Schwarzschild-de Sitter solution, $\Lambda
>0$, we are only interested in the range $r_b<a_0<r_c$, with $r_b$
and $r_c$ given by equation (\ref{root1}) and equation
(\ref{root2}), respectively. The factor
$g(a_0)=(1-3M/a_0+3M^2/a_0^2-\Lambda Ma_0)$ has only one positive
root, $a_r$, so that $g(a_0)>0$ for $a_0 < a_r$, and $g(a_0)<0$
for $a_0
> a_r$. It is a simple matter to prove that $a_r>r_c$, i.e., the
value for which the factor $g(a_0)$ becomes negative is greater
than the cosmological horizon, $r_c$. In particular, if $\Lambda
M^2 \ll 1$, we verify that $a_r\gg r_c$. For transparency,
consider the following algebraic analysis, analogous to the one
preceding Figure $1$. Consider the factor
$f(a_0)=(1-2M/a_0-\Lambda a_0^2/3)$. Define $\alpha =2M/a_0$ and
$\beta = 9 \Lambda M^2$, so that $\beta =27\alpha^2(1-\alpha)/4$
is considered for $f(a_0)=0$, depicted as a solid line in Figure
$2$. $g(a_0)=0$ or $9\Lambda M^2 =2\alpha(1-3\alpha/2+3\alpha
^2/4)/9$ is shown as a dashed line in Figure $2$. The region of
values for which $g(a_0)>0$ is depicted below the dashed line,
which covers the entire area of interest. Thus, the
right-hand-side of equation (\ref{inequality}) in the range of
interest, $r_b<a_0<r_c$, is strictly negative.

\begin{figure}[h]
  \centering
  \includegraphics[width=2.8in]{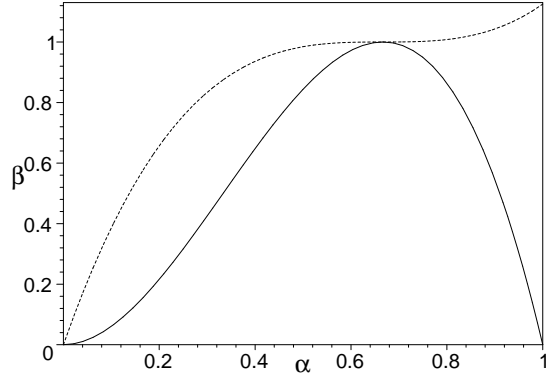}
  \caption{We have defined $\alpha=2M/a_0$ and $\beta=9\Lambda M^2$.
  Only the region below the solid line is considered. The region of values for
  which $g(a_0)>0$ is depicted below the dashed line. See text for
  details.}
\end{figure}

Secondly, for the Schwarzschild-anti de Sitter spacetime, $\Lambda
<0$, we verify that the factor
$h(a_0)=(1-3M/a_0+3M^2/a_0^2+|\Lambda |Ma_0)$ is always positive.

Thirdly, if $M=0$ and $\Lambda \neq 0$, the solution reduces to
the de Sitter spacetime, if $\Lambda > 0$, and to the anti-de
Sitter, if $\Lambda < 0$, we verify that the right-hand-side of
equation (\ref{inequality}) is always negative. If $\Lambda = 0$
and $M \neq 0$, the analysis is reduced to the Schwarzschild
solution, considered by Poisson and Visser \cite{Poisson}.

Therefore, for all the cases considered above, the right hand side
of equation (\ref{inequality}) is always negative, while the left
hand side changes sign at $a_0=3M$. Thus, one deduces that the
stability regions are dictated by the following inequalities
\begin{eqnarray}
\eta_0<-\frac{1-3M/a_0+3M^2/a_0^2-\Lambda
Ma_0}{2\left(1-2M/a_0-\Lambda a_0^2/3 \right)\left(1-3M/a_0
\right)}\,,& \,a_0>3M
    \\
\eta_0>-\frac{1-3M/a_0+3M^2/a_0^2-\Lambda
Ma_0}{2\left(1-2M/a_0-\Lambda a_0^2/3 \right)\left(1-3M/a_0
\right)}\,,\qquad& \,a_0<3M .
\end{eqnarray}

One may analyze several cases.

\subsubsection{De Sitter spacetime.}

For the de Sitter spacetime, with $M=0$ and $\Lambda >0$, equation
(\ref{inequality}) reduces to
\begin{equation}
\eta_0<-\frac{1}{2\left(1-\Lambda a_0^2/3 \right)} \,, \;\;\;\;
{\rm for}\;\;0<a_0<\sqrt{3/\Lambda} \,.
\end{equation}
The stability region is depicted in the left plot of figure $3$.

\subsubsection{Anti-de Sitter spacetime.}

For the anti-de Sitter spacetime, with $M=0$ and $\Lambda <0$,
equation (\ref{inequality}) gives
\begin{equation}
\eta_0<-\frac{1}{2\left(1+|\Lambda| \,a_0^2/3 \right)} \,,
\;\;\;\; {\rm for}\;\;a_0>0   \label{adsconstraint}\,,
\end{equation}
and the respective stability region is depicted in the right plot
of figure $3$. As mentioned in the Introduction the role of a
negative cosmological constant plays an important role in the
Randall-Sundrum models \cite{Randall,Randall2}. Anchordoqui and
Olsen \cite{Anch2} considered the dynamics of a $d$-dimensional
brane-world sweeping through the $(d+1)$-dimensional bulk, by
imposing constraints on the brane-world cosmologies and
considering that $V''(a_0)<0$, so that $V(a_0)$ is a local maximum
at $a_0$. In particular, they analyzed different behaviours for
the expansion of the universe by imposing constraints on the speed
of sound. But, in the presence of exotic matter one cannot naively
interpret $\eta_0$ as the speed of sound on the wormhole throat.
Therefore the plots that we have considered generally extend
beyond $0<\eta_0 \leq 1$. Equation (\ref{adsconstraint}) may be
readily obtained from the Anchordoqui-Olsen linearized stability
analysis by setting $d=3$ and considering that $V(a_0)$ is a local
minimum so that $V''(a_0)>0$.

\begin{figure}[h]

  \centering
  \includegraphics[width=2.4in]{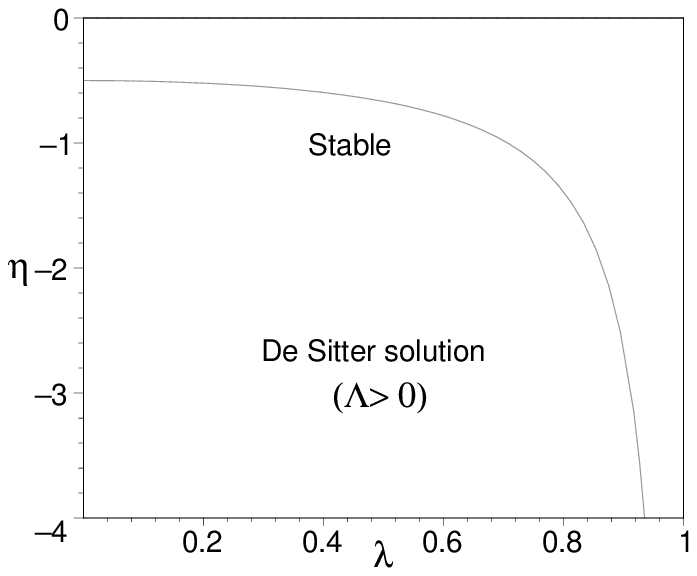}
  \hspace{0.4in}
  \includegraphics[width=2.4in]{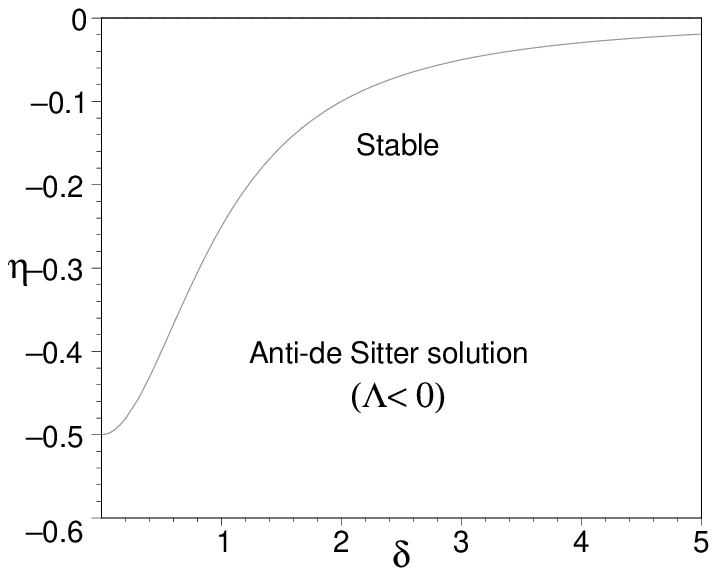}
  \caption{We have defined $\lambda = a_0/(3/\Lambda)^{1/2}$ and
  $\delta = a_0/(3/|\Lambda|)^{1/2}$, respectively.
  The regions of stability are depicted in the graphs,
  below the curves for the de-Sitter and the anti-de Sitter solutions, respectively.}
\end{figure}

\subsubsection{Schwarzschild spacetime.}

This is the particular case of the Poisson-Visser analysis
\cite{Poisson}, with $\Lambda=0$, which reduces to
\begin{eqnarray}
\eta_0<-\frac{1-3M/a_0+3M^2/a_0^2}{2\left(1-2M/a
\right)\left(1-3M/a_0 \right)}\,, & a_0>3M \,,  \\
\eta_0>-\frac{1-3M/a_0+3M^2/a_0^2}{2\left(1-2M/a
\right)\left(1-3M/a_0 \right)}\,, \qquad& a_0<3M \,.
\end{eqnarray}
The stability regions are shown in the left plot of figure $4$.

\subsubsection{Schwarzschild-anti de Sitter spacetime.}

For the Schwarzschild-anti de Sitter spacetime, with $\Lambda <0$,
we have
\begin{eqnarray}
\eta_0<-\frac{1-3M/a_0+3M^2/a_0^2+|\Lambda| \,
Ma_0}{2\left(1-2M/a_0+|\Lambda|\, a_0^2/3 \right)\left(1-3M/a_0
\right)}  \,, & a_0>3M     \label{ads1}
    \\
\eta_0>-\frac{1-3M/a_0+3M^2/a_0^2+|\Lambda| \,
Ma_0}{2\left(1-2M/a_0+|\Lambda|\, a_0^2/3 \right)\left(1-3M/a_0
\right)}\,, \qquad& a_0<3M        \label{ads2}
\end{eqnarray}

The regions of stability are depicted in the right plot of figure
$4$, considering the value $9|\Lambda |M^2=0.9$. In this case, the
black hole event horizon is given by $r_b \simeq 1.8\,M$. We
verify that the regions of stability decrease, relatively to the
Schwarzschild case.

Various aspects of the five dimensional Schwarzschild-anti de
Sitter solution were treated in \cite{Kraus,Chamblin}. Kraus
considered the dynamics of anti-de Sitter domain wall with a
static bulk \cite{Kraus}, by deducing equations of motion of the
wall applying the Darmois-Israel formalism. Observers on the wall
interpret the motion of the wall through the static background as
a cosmological expansion or contraction. One may generalize the
$(d+1)$-dimensional anti-de Sitter Anchordoqui-Olsen linearized
stability analysis \cite{Anch2} to the $(d+1)$-dimensional
Schwarzschild-anti de Sitter solution, and obtain equations
(\ref{ads1})-(\ref{ads2}) by setting $d=3$.

\begin{figure}[h]
  \centering
  \includegraphics[width=2.4in]{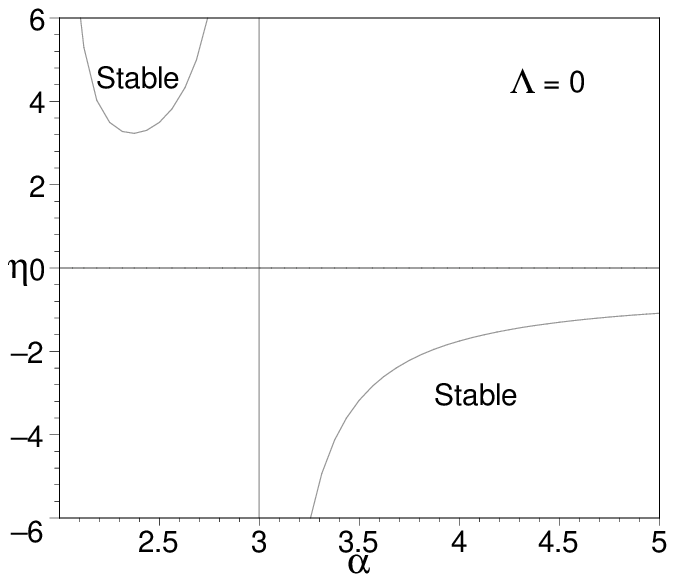}
  \hspace{0.4in}
  \includegraphics[width=2.4in]{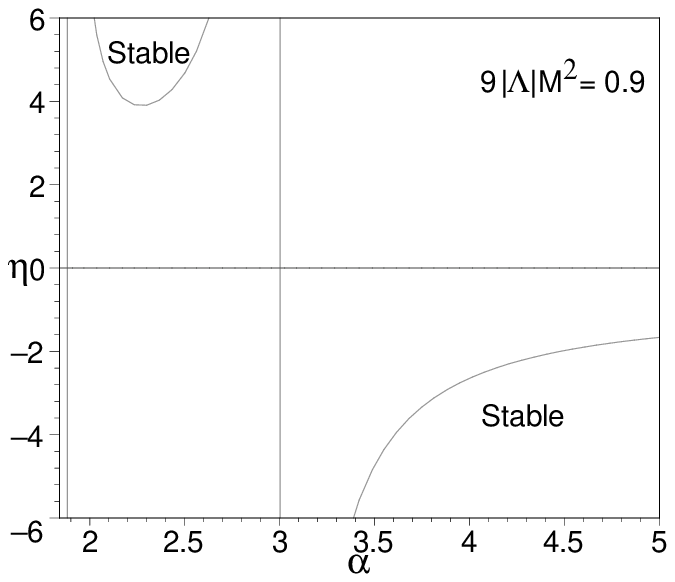}
  \caption{We have defined $\alpha =a_0/M$.
  The regions of stability are depicted for the
  Schwarzschild and the Schwarzschild-anti de Sitter solutions, respectively.
  Imposing the value $9|\Lambda |M^2=0.9$ in the Schwarzschild-anti de Sitter case,
  we verify that the stability regions decrease relatively to the
  Schwarzschild solution.}
\end{figure}

\subsubsection{Schwarzschild-de Sitter spacetime.}

For the Schwarzschild-de Sitter spacetime, with $\Lambda
>0$, we have
\begin{eqnarray}
\eta_0<-\frac{1-3M/a_0+3M^2/a_0^2-\Lambda
Ma_0}{2\left(1-2M/a_0-\Lambda a_0^2/3 \right)\left(1-3M/a_0
\right)}\,, & \,a_0>3M
    \\
\eta_0>-\frac{1-3M/a_0+3M^2/a_0^2-\Lambda
Ma_0}{2\left(1-2M/a_0-\Lambda a_0^2/3 \right)\left(1-3M/a_0
\right)}\,, \qquad& \,a_0<3M .
\end{eqnarray}

The regions of stability are depicted in figure $5$ for increasing
values of $9\Lambda M^2$. In particular, for $9\Lambda M^2=0.7$
the black hole and cosmological horizons are given by $r_b \simeq
2.33\,M$ and $r_c =4.71\,M$, respectively. Thus, only the interval
$2.33 < a_0/M < 4.71$ is taken into account, as shown in the range
of the respective plot. Analogously, for $9\Lambda M^2=0.9$, we
find $r_b \simeq 2.56\,M$ and $r_c =3.73\,M$. Therefore only the
range within the interval $2.56 < a_0/M < 3.73$ corresponds to the
stability regions, also shown in the respective plot.

We verify that for large values of $\Lambda$, or large $M$, the
regions of stability are significantly increased, relatively to
the $\Lambda =0$ case.

\begin{figure}[h]
\centering
  \includegraphics[width=2.4in]{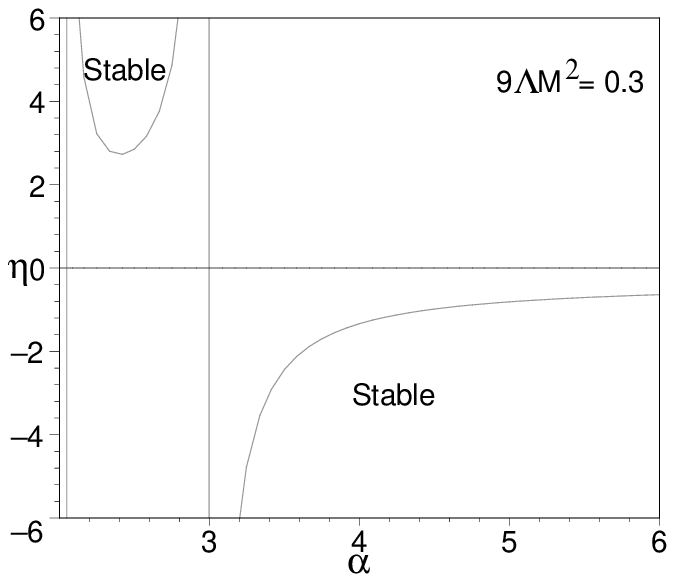}
  \hspace{0.4in}
  \includegraphics[width=2.4in]{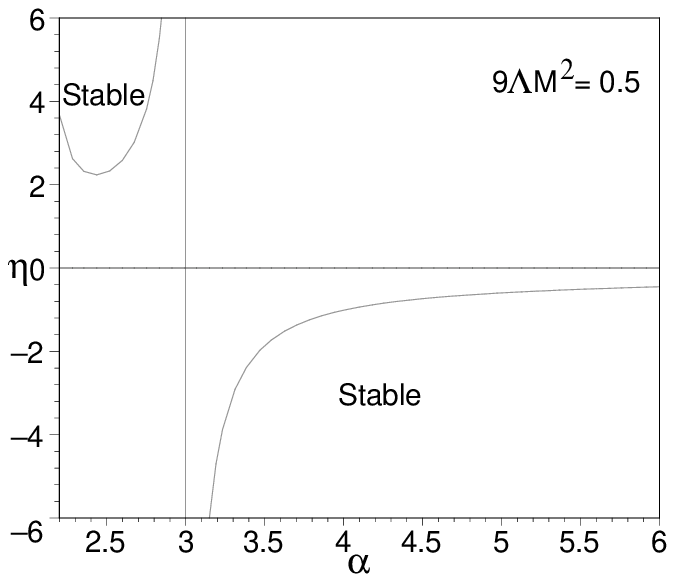}
  \centering
  \includegraphics[width=2.4in]{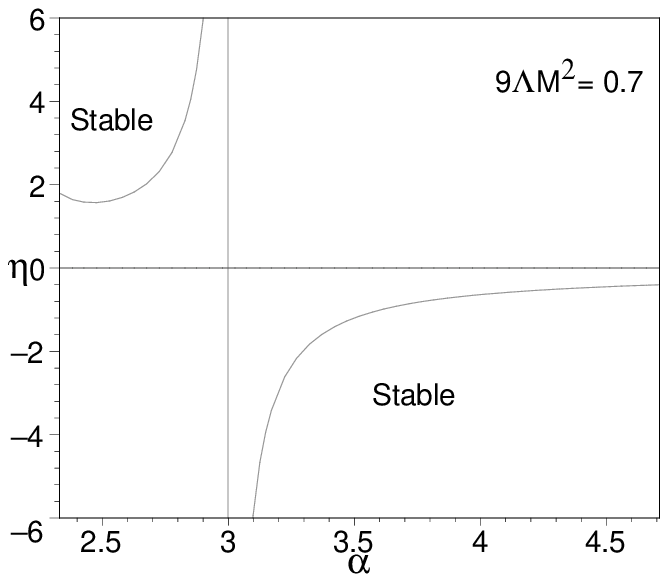}
  \hspace{0.4in}
  \includegraphics[width=2.4in]{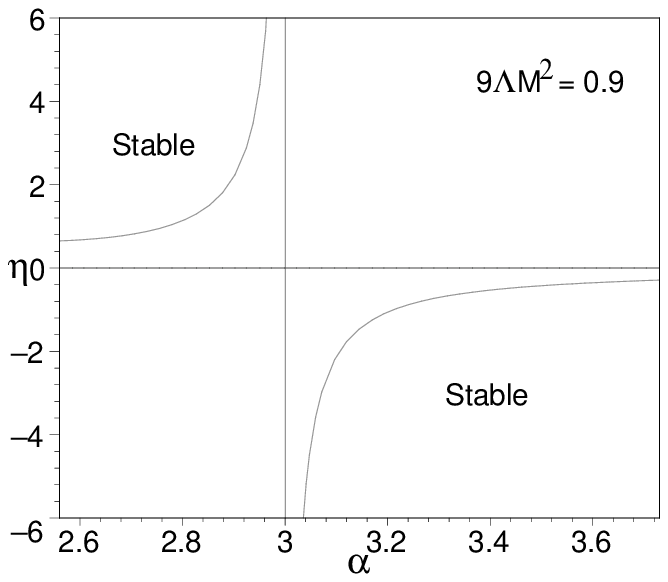}
  \caption{We have defined $\alpha =a_0/M$.
  The regions of stability for the Schwarzschild-de Sitter solution,
  imposing $9\Lambda M^2=0.3$, $9\Lambda M^2=0.5$, $9\Lambda M^2=0.7$
  and $9\Lambda M^2=0.9$, respectively.
  The regions of stability are significantly increased, relatively
  to the $\Lambda =0$ case, for increasing values of $9\Lambda M^2$.}
\end{figure}

\section{Conclusion}

The traditional manner of solving the Einstein field equation
consists in considering a plausible stress-energy tensor and
finding the respective geometrical structure. Specifically,
physically plausible equations of state are derived for the
respective pressures as a function of the energy density, and
using the Einstein field equation, the respective metric is
determined. In considering wormhole solutions, the Einstein field
equation is run in the reverse direction by imposing an exotic
metric, $g_{\mu \nu}$, and eventually the matter source for the
respective geometry is found. In this manner, solutions involving
the violation of the energy conditions have been found. However,
up to date a detailed microphysical model describing the physics
of exotic matter does not exist, i.e., nothing is known on
equations of state of exotic matter. Under normal circumstances
the parameter $\sqrt{\eta_0}$ is interpreted as the speed of
sound, but in the presence of exotic matter this cannot naively be
done so. Consequently, one cannot {\it a priori} impose $0 <
\eta_0 \leq 1$, as there are several known examples of exotic
$\eta_0 <0$ behavior, namely the Casimir effect and the false
vacuum. See \cite{Poisson} for a detailed analysis. Therefore, the
conditions $\eta_0 >1$ and $\eta_0 < 0$ cannot be ruled out, until
a detailed microphysical model of exotic matter is devised.

However, we have found that for large positive values of
$\Lambda$, i.e., the Schwarzschild-de Sitter solution, the regions
of stability significantly increase relatively to the case
analyzed by Poisson and Visser, imposing less severe restrictions
on $\eta_0$. For negative values of $\Lambda$, i.e., the
Schwarzschild-anti de Sitter, the regions of stability decrease.



\appendix

\section{Linearized stability in $(2+1)$-dimensional
Schwarzschild-de Sitter thin-shell wormholes}

It is interesting to analyze the $(2+1)$-dimensional
Schwarzschild-de Sitter solution \cite{Kim2}. The metric is given
by
\begin{equation}
ds^2=-\frac{(1-4M-\Lambda\,r^2)}{1-4M}
\,dt^2+\frac{dr^2}{1-4M-\Lambda\,r^2}+r^2\, d\phi ^2 \label{SdS3}
\,.
\end{equation}
For $1-4M \geq 0$, an event horizon appears at
$r_b=\sqrt{(1-4M)/\Lambda}$. We shall only consider the static
spacetime, with $r<r_b$.

If $\Lambda =0$, eq. \eref{SdS3} is reduced to the
$(2+1)$-dimensional Schwarzschild metric,
\begin{equation}
ds^2=-dt^2+\frac{dr^2}{1-4M}+r^2\, d\phi ^2 \label{Schw3} \,.
\end{equation}
The properties and characteristics of the metric \eref{Schw3} are
discussed in \cite{Kim2,Perry}.

Considering $M=0$, eq. \eref{SdS3} reduces to the
$(2+1)$-dimensional de Sitter metric, given by
\begin{equation}
ds^2=-(1-\Lambda\,r^2) \,dt^2+\frac{dr^2}{1-\Lambda\,r^2}+r^2\,
d\phi ^2 \label{dS3} \,.
\end{equation}

Considering the cut-and-paste construction described in section
$2.3$, the Lanczos equations in $(2+1)$-dimensions reduce to
\begin{eqnarray}
\sigma &=&-\frac{1}{8\pi}\kappa ^{\theta}_{\;\,\theta} \,,   \label{3sigma} \\
p &=&+\frac{1}{8\pi}\kappa ^{\tau}_{\;\,\tau} \label{linepressure}
\,.
\end{eqnarray}
The extrinsic curvatures are given by
\begin{eqnarray}
K^{\theta\;\pm}_{\;\,\theta}&=& \pm \frac{1}{a}\sqrt{1-4M-\Lambda
a^2+\dot{a}^2} \;,
    \label{3Kplustheta}\\
K ^{\tau \;\pm}_{\;\,\tau}&=& \pm \frac{-\Lambda a+ \ddot{a}}
{\sqrt{1-4M-\Lambda a^2+\dot{a}^2}} \;,
    \label{3Kplustautau}
\end{eqnarray}
and eqs. \eref{3sigma}-\eref{linepressure} take the following form
\begin{eqnarray}
\sigma&=&-\frac{1}{4\pi a} \sqrt{1-4M-\Lambda a^2+\dot{a}^2}
  \label{3surfenergy}   \,,\\
p&=&-\frac{1}{4\pi a} \;\frac{\Lambda a^2-a\ddot{a}
}{\sqrt{1-4M-\Lambda a^2+\dot{a}^2}} \label{3surfpressure}  \,.
\end{eqnarray}

The conservation of the line stress-energy tensor implies
\begin{equation}
\dot{\sigma}=-\left(\sigma +p \right) \frac{\dot{a}}{a}
   \label{3conservationenergy}
\end{equation}
or
\begin{equation}
\frac{d\left(\sigma l \right)}{d\tau}+p\,\frac{dl}{d\tau}=0 \,,
\end{equation}
where $l=2\pi a$ is the circumference of the stress-energy ring .

The equation of motion of the wormhole throat, deduced from eq.
\eref{3surfenergy}, takes the following form
\begin{equation}
\dot{a}^2-4M-\Lambda a^2- \left(4\pi a \sigma \right)^2=-1 \,,
\label{3motionofthroat}
\end{equation}
or by $\dot{a}^2=-V(a)$, with the potential defined as
\begin{equation}
V(a)=1-4M-\Lambda a^2- \left(4\pi a \sigma \right)^2
\label{3defpotential}  \,.
\end{equation}

Linearizing around the stable solution at $a=a_0$, consider the
Taylor expansion of $V(a)$ around $a_0$ to second order in section
$3.2$. The static line energy density and line pressure are given
by
\begin{eqnarray}
\sigma_0&=&-\frac{1}{4\pi a_0} \sqrt{1-4M-\Lambda
a_0^2}   \label{3stablesigma} \,,\\
p_0&=&-\frac{1}{4\pi a_0}\;\frac{\Lambda a_0^2}{\sqrt{1-4M-\Lambda
a_0^2}} \,. \label{3stablepressure}
\end{eqnarray}
Considering eq. \eref{Taylorexpansion}, the first and second
derivatives of $V(a)$, are
\begin{eqnarray}
V'(a)&=&-2\Lambda a+32\pi^2 a \sigma p   \label{3-firstderivativeV}  \,, \\
V''(a)&=&-2\Lambda-32\pi^2 \big[p^2 +\sigma (\sigma +p)\eta \big]
\label{3-2ndderivative} \,,
\end{eqnarray}
taking into account the definition $\eta(\sigma)=p'/\sigma'$ and
rewriting the conservation of the line stress-energy tensor,
equation (\ref{3conservationenergy}), as $\sigma 'a=-(\sigma +p)$,
with $\sigma '=\dot{\sigma}/\dot{a}$.

Evaluated at the static solution, at $a=a_0$, we readily find
$V(a_0)=0$ and $V'(a_0)=0$, with $V''(a_0)$ given by
\begin{equation}
V''(a_0)=-\frac{2}{a_0^2}\, (1-4M) \left(\frac{\Lambda
a_0^2}{1-4M-\Lambda a_0^2}+\eta_0 \right)\,.
\end{equation}

The solution is stable if and only if $V''(a_0)>0$, or
\begin{equation}
\eta_0 < -\frac{\Lambda a_0^2}{1-4M-\Lambda a_0^2}
   \label{3inequality}   \,.
\end{equation}


\end{document}